\documentclass[12pt,tightenlines,nofootinbib,preprintnumbers,amsmath,amssymb]{revtex4}

\usepackage{hyperref}

\date{March 2010}
\usepackage{epsfig}
\usepackage{latexsym}
\usepackage{amssymb}
\usepackage{booktabs}
\usepackage{graphicx}
\newcommand{\be}{\begin{equation}}
\newcommand{\ee}{\end{equation}}
\newcommand{\ba}{\begin{eqnarray}}
\newcommand{\ea}{\end{eqnarray}}
\newcommand{\bi}{\begin{itemize}}
\newcommand{\ei}{\end{itemize}}

\newcommand{\<}{\langle}
\renewcommand{\>}{\rangle}

\newcommand{\la}{\label}

\newcommand{\txts}{\textstyle}

\begin{document}
\preprint{MITP-23-056}
\title{Low-energy matrix elements of heavy-quark currents}

\author{Harvey~B.~Meyer}
 \affiliation{PRISMA$^+$ Cluster of Excellence  \& Institut f\"ur Kernphysik,
Johannes Gutenberg-Universit\"at Mainz, D-55099 Mainz, Germany}
\affiliation{Helmholtz-Institut Mainz, Johannes Gutenberg-Universit\"at Mainz,
D-55099 Mainz, Germany}

\date{\today}

\begin{abstract}
In QCD at energies well below a heavy-quark threshold, the heavy-quark vector current
can be represented via local operators made of the lighter quarks and of the gluon fields.
We extract the leading perturbative matching coefficients for the two most important sets of operators from known results.
As an application, we analytically determine the  ${\rm O}(\alpha_s^3)m_c^2/m_b^2$ effect of the bottom quark current on the $R(s)$ ratio
below the bottom but above the charm threshold. 
For the low-energy representation of the charm quark current, 
the two most important operators are given by the total divergence of dimension-six gluonic operators.
We argue that the charm magnetic moment of the nucleon is effectively measuring the forward matrix elements
of these gluonic operators and predict the corresponding bottom magnetic moment. Similarly, the contribution of the charm current
to $R(s\approx 1\,{\rm GeV}^2)$, which is associated with quark-disconnected diagrams, is dominantly determined by the decay constants
of the $\omega$ and $\phi$ mesons with respect to the two gluonic operators.
\end{abstract}

\maketitle

\section{Introduction}

Imagine a world of strong and electromagnetic interactions in which the up, down and strange quarks
are electrically neutral.
What would be the size of the magnetic moment of the proton?
With what signal strength would the $\omega$ and $\phi$ mesons show up in the famous ratio of cross-sections
\be
R(s) = \frac{\sigma(e^+e^- \rightarrow {\rm hadrons})}{\sigma(e^+e^- \rightarrow \mu^+\mu^-)} \qquad {\rm ?}
\ee
More generally, what form would $R(s)$ take below the $J/\psi$ resonance? These are the sorts of questions we are after in this paper.

These questions may at first seem to be of purely academic interest.
However, there are a few reasons to pay attention to them.  One is
that the observables in the aforementioned world can be isolated and
computed rigorously in lattice QCD, since it is straightforward in
that framework to keep the $(u,d,s)$ electric charges `switched off'.
In particular, a first calculation of the nucleon charm magnetic
moment has appeared~\cite{Sufian:2020coz}, finding a negative value on
the order of $10^{-3}$ in units of the nuclear magneton. In the
context of high-precision calculations of the hadronic vacuum
polarisation (HVP), the sub-threshold effects due to the heavy quarks
are associated with the quark-disconnected diagrams involving at least
one heavy-quark loop.  The charm disconnected diagrams have been
reported to be very small~\cite{Budapest-Marseille-Wuppertal:2017okr}
in a calculation of the hadronic vacuum polarisation contribution to
the muon $(g-2)$ performed on a coarse lattice: less than one
percent of the $(u,d,s)$ disconnected contribution.  However, 
the charm disconnected loops are bound to be less suppressed
at somewhat higher virtualities, in the vacuum polarisation or in the
closely related running of the weak mixing angle~\cite{Jegerlehner:1985gq,Ce:2022eix}.
In any case, theoretical predictions for the charm-quark contributions can be
compared unambiguously with lattice QCD calculations.

A second motivation is that in the process of deriving these
theoretical predictions, one gains insight into what matrix elements
in the low-energy effective theory the heavy-quark contribution is
really picking up. By the same token, the predictions can fairly
straightforwardly be extended to the bottom quark, which is not easily
handled dynamically in lattice QCD due to its large mass.

A third, much longer-term motivation, is that the proton magnetic
moment $\mu_p$ is known experimentally to a precision of
0.3\,ppb~\cite{Tiesinga:2021myr}. In principle, it could be used to
search for new physics, as is done for the anomalous magnetic moment
of the muon $a_\mu = (g-2)_\mu/2$, if the Standard Model prediction
could be made significantly more precise.  The generic sensitivity to
heavy degrees of freedom is enhanced due to the
heavier mass scale of the proton.  Since its magnetic moment is a
non-perturbative quantity from the outset, and not just starting at
O($\alpha^2$) as for $a_\mu$, the hadronic uncertainties enter at
O(1), making the theoretical task enormously harder. For instance, a
recent lattice QCD calculation~\cite{Djukanovic:2023jag} achieved a
precision of 2.4\% on the magnetic moment of the proton. Nevertheless,
from this point of view one might be curious to know whether the current
experimental precision on the proton magnetic moment already makes it
sensitive to the bottom, or even to the top quark contributions to the
electromagnetic current.

The questions formulated in the introductory paragraph can be
addressed by `integrating out' the heavy quark.  In a low-energy
matrix element, the heavy quark-antiquark pair coupling to the
external photon annihilates in a short-distance process. Via a
three-gluon intermediate state, it can act as a light-quark bilinear.
The total divergence of the antisymmetric tensor
current is the lowest-dimensional operator that has the right symmetry
properties. However, it is a helicity-flip operator, and therefore
cannot contribute if the light quarks are actually massless. This
leads to an additional factor of the light-quark masses appearing in
this operator, making it effectively of dimension five, and therefore
suppressed by $1/m_Q^2$. This operator is considered in section
\ref{sec:tens_curr}.  Looking for further, not chirally suppressed contributions,
we note that it is not possible to construct a low-energy
effective operator with just two field strength tensors, leading us
to consider (dimension-seven) operators built out of three field
strength tensors in section \ref{sec:EH}.  Operators such as 
$m_f\partial_\nu(\bar\psi_f G_{\mu\nu}\psi_f)$
are of the same dimension, but chirally suppressed, and we will therefore not consider
them.

Thus the low-energy representation of the heavy-quark vector current is power-suppressed.
In this paper we will therefore be discussing effects that are very small, but this must be weighed against the fact that 
matrix elements of the electromagnetic current are known to very high precision
and are continuously being improved upon.
It is worth contrasting the low-energy representation of a heavy-quark \emph{vector} current with that of the corresponding \emph{axial-vector} current,
which has been worked out long ago~\cite{Collins:1978wz}. The leading result for the mapping between
renormalisation-group invariant currents reads~\cite{Chetyrkin:1993hk}
\be \la{eq:axialLE}
(\bar Q \gamma^\mu\gamma^5 Q)_{\rm RGI} \stackrel{m_Q\to\infty}{\longrightarrow}
 - \frac{6}{33-2(N_f+1)} \Big(\frac{\alpha_s(m_Q)}{\pi}\Big) \sum_{f=1}^{N_f} (\bar q_f \gamma^\mu\gamma^5 q_f)_{\rm RGI}.
\ee
That is, the low-energy matrix elements of the axial current of a heavy quark are asymptotically suppressed by $1/\log(m_Q^2/\Lambda_{\rm QCD}^2)$,
as opposed to a power law. A further example is the Lagrangian mass term of a heavy quark, $m_Q\bar Q Q$, which does not decouple
as $m_Q\to\infty$, but rather contributes to the trace anomaly of the low-energy effective theory~\cite{Shifman:1978zn}.

In section~\ref{sec:tens_curr}, we work out the matching coefficient of the light-quark tensor currents
to the heavy-quark vector current.
In section~\ref{sec:EH}, we work out the matching coefficient of the three-gluon operators
to the heavy-quark vector current.
We turn to physics applications in section \ref{sec:appl} and conclude in section \ref{sec:concl}.

\section{The tensor currents of the light quarks \la{sec:tens_curr}}

\begin{figure}
  \centerline{\includegraphics[width=0.18\textwidth]{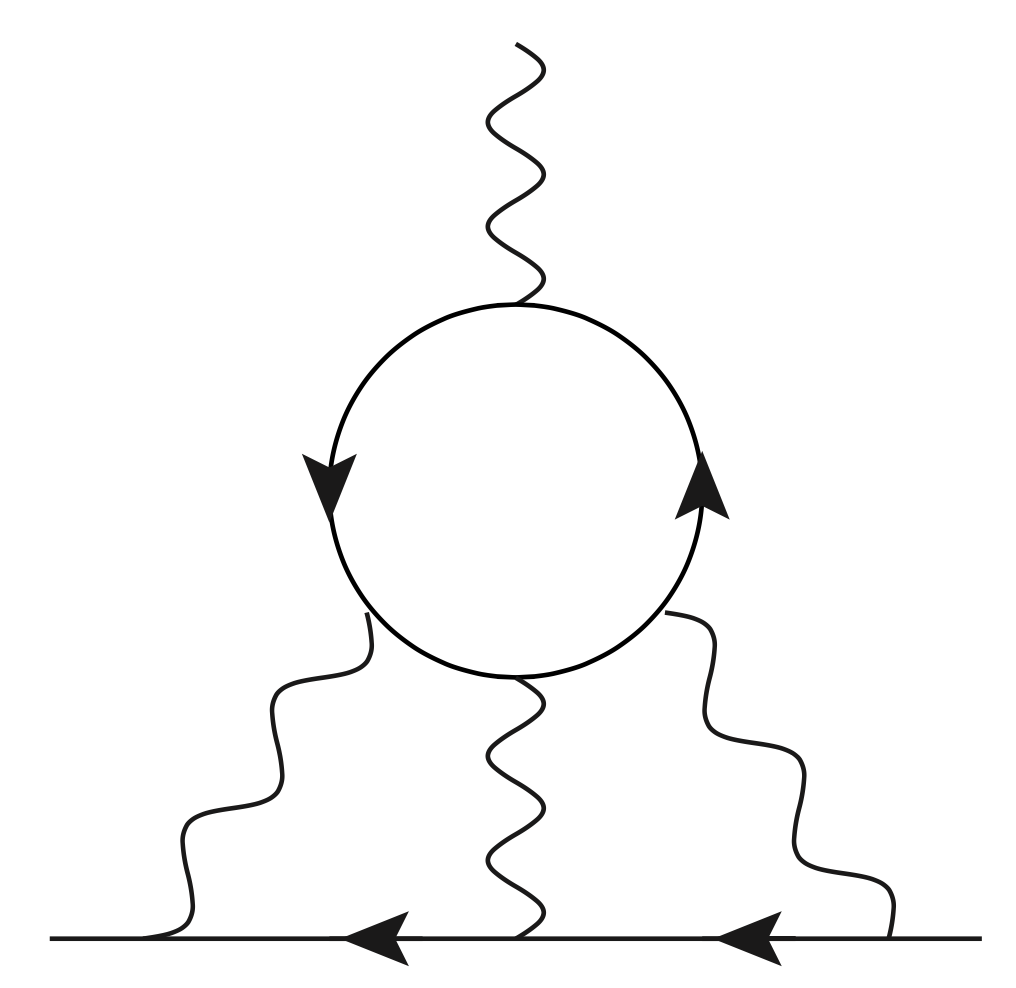}~~~~~~~~~~~~~~~~~~~~
    \includegraphics[width=0.25\textwidth]{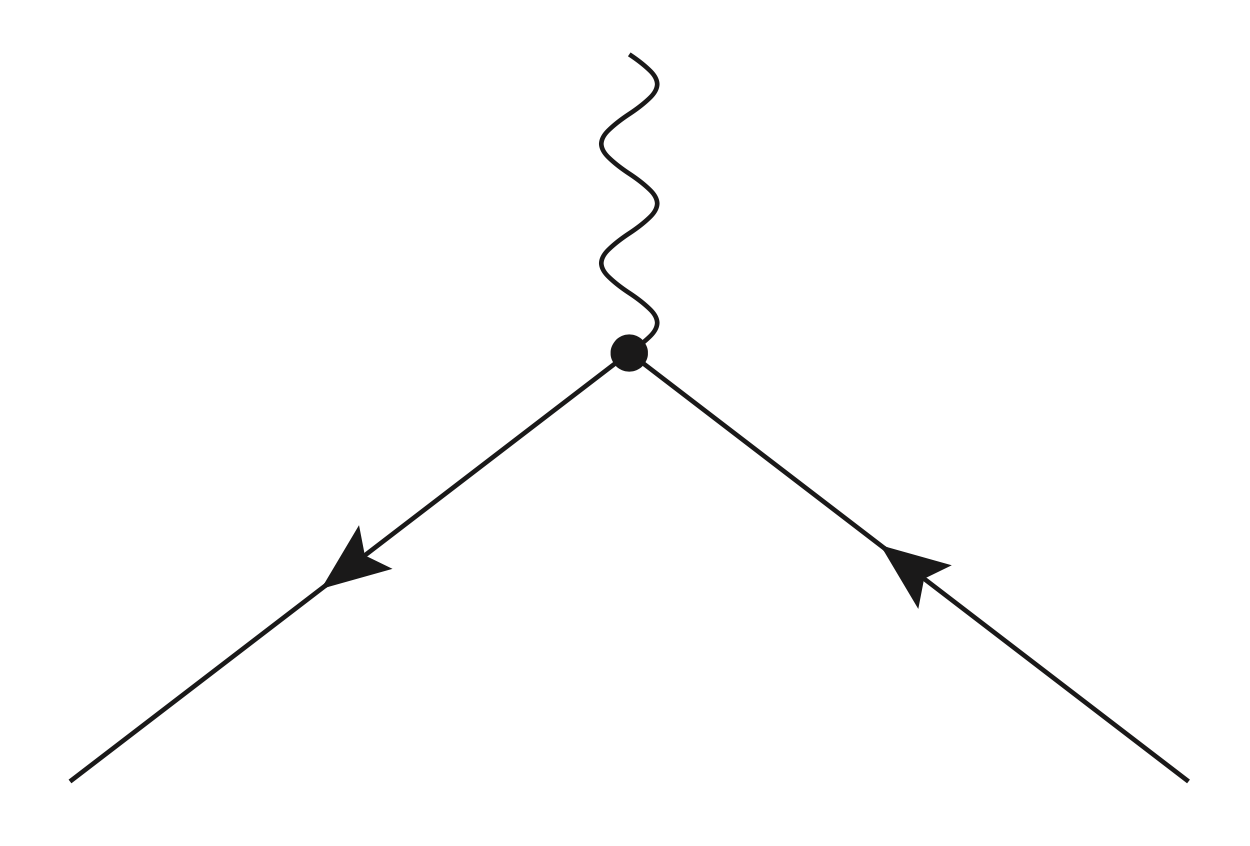}}
  \caption{\la{fig:lblgm2} Left: representative diagram of the light-by-light contribution to the $g-2$ of an electron via a muon loop.
    Right: Effective interaction between the electron and the photon induced by the muon loop, described by the total divergence of the antisymmetric
    tensor current.
  }
  \end{figure}
Consider QED with two types of leptons $\ell$ and $L$, respectively of masses $m$ and $M$.
The latter is assumed to be far greater than the former, $m\ll M$, and we denote the action by $S$.
At scales well below $M$, and
to leading order, the effective field theory is QED with the lepton $\ell$ only, which has the standard action $S_0$.
We refer the reader to the lecture notes~\cite{Grozin:2009an} for more details.
But what are the low-energy matrix elements of the heavy-lepton current $\bar L\gamma^\mu L$?
This question can be answered 
by minimally coupling the heavy-lepton current
to an external electromagnetic field $B_\mu(x)$, so that $S\to S -e_L \int d^4x\, B_\mu \bar L \gamma^\mu L$, the constant $e_L$
representing the charge of the heavy lepton.

Correspondingly, the low-energy effective action is extended from $S_0$ to $(S_0+ S^{\rm ext})$.
The generating functional $F[B]$ describing the complete theory in the presence of the external field 
is approximated at low energies by the generating functional $F_{0}[B]$ given by
$ \exp(iF_{0}[B]) = \<0| {\rm T}\;\exp (iS^{\rm ext}) |0\>$,
where the dynamics in the latter matrix element are governed by the action $S_0$.
The connected correlation functions of the heavy-quark current are then obtained by taking functional derivatives
of $F[B]$ at $B_\mu=0$, respectively of $F_{0}[B]$ in the effective theory.
The leading form of $S^{\rm ext}$ in powers of $1/M$ is given by the Pauli term\footnote{Note that we use a Minkowski-space notation with `mostly minus' metric
and $\{\gamma^\mu,\gamma^\nu\} = 2\eta^{\mu\nu}$. A transcription into Euclidean notation is given at the beginning of section \ref{sec:appl}.},
\ba\la{eq:Sext2l1}
S^{\rm ext} &=& e_L \cdot d_2^{\rm QED}\,\int d^4x\; (\partial_\mu B_\nu - \partial_\nu  B_\mu)\, \bar \ell {\txts\frac{i}{2}}[\gamma^\mu,\gamma^\nu]\ell
\\ &=& -e_L\cdot  d_2^{\rm QED}\,\int d^4x\;   B_\mu\, (-i) \partial_\nu(\bar \ell [\gamma^\mu,\gamma^\nu]\ell),
\la{eq:Sext2l2}
\ea
with $d_2^{\rm QED}$ a matching coefficient to be determined below.
If we are only interested in those correlation functions of the current $\bar L \gamma^\mu L$
with a finite separation in position space between the current insertions (which is the case throughout this paper), 
the use of the generating function is equivalent to substituting each current by the divergence of the antisymmetric tensor
current, as it appears in Eq.\ (\ref{eq:Sext2l2}). Indeed, $-i\partial_\nu(\bar \ell [\gamma^\mu,\gamma^\nu] \ell)$
plays the role of  a conserved current with vanishing forward matrix elements on a light-lepton state.
In the chiral limit, this operator is however chirality-flipping, which $\bar L\gamma^\mu L$ is not;
therefore, we can anticipate that our operator will be accompanied by a factor $m$.

By dimensional analysis, the matching coefficient $d_2^{\rm QED}$ must be of order $1/M^2$.
It can be obtained from the following consideration (Fig.\ \ref{fig:lblgm2}).
The contribution to the anomalous magnetic moment of lepton $\ell$ of the current
$-i\partial_\nu(\bar \ell [\gamma^\mu,\gamma^\nu] \ell)$ is
\be
\Delta F_2(q^2=0) = - 4m,
\ee
while the contribution to $F_1(0)$ vanishes.
On the other hand, the direct (three-loop) calculation of the anomalous magnetic moment
contribution of the current $\bar L\gamma^\mu L$ yields~\cite{Laporta:1992pa,Kuhn:2003pu}
\ba
\Delta F_2(0) &=& c_2 \left(\frac{\alpha}{\pi}\right)^3\,\left(\frac{m}{M}\right)^2 + {\rm O}((m/M)^4),
\\
c_2 &=&  \frac{3}{2} \zeta(3) - \frac{19}{16}.
\ea
Hence we have the following mapping
\ba\la{eq:mappingQEDtensorcurr}
&& \bar L \gamma^\mu L \longrightarrow d_2^{\rm QED}(M,m,\alpha) (-i)\partial_\nu(\bar\ell [\gamma^\mu,\gamma^\nu]\ell),
\\ && d_2^{\rm QED}(M,m,\alpha) =  -\frac{c_2}{4} \frac{m}{M^2} \,\left(\frac{\alpha}{\pi}\right)^3,
\ea
of the current in the complete theory into the low-energy EFT.
The two-point function
\ba\la{eq:mixed2ptQED}
&&  f^{\mu\nu}(x;M,m) \equiv \< \bar L(x) \gamma^\mu L(x)\; \bar\ell(0)\gamma^\nu\ell(0)\>
\\ &&\to 
d_2^{\rm QED}(M,m,\alpha) \< (-i)\partial_\lambda (\bar\ell(x) [\gamma^\mu,\gamma^\lambda]\ell(x))\; \bar\ell(0) \gamma^\nu \ell(0)\>
\phantom{\Big|}
\nonumber
\ea
can thus be evaluated within QED with a single light lepton for large spacelike separations, $-M^2x^2 \gg 1$.

Similar considerations apply to QCD with $N_f$ `light' quark flavours
and one additional heavy flavour $Q$. The heavy quark current $\bar Q\gamma_\mu Q$ can be matched
to an operator made of fields in the `light' sector.
The matching coefficient is the same
as in QED, up to a colour factor. To determine this factor, consider the two-point function
\be\la{eq:mixed2ptQCD}
\< \bar Q(x) \gamma^\mu Q(x)\; \bar q(0)\gamma^\nu q(0)\> = \frac{d^{abc}}{4} \frac{d^{abc}}{4} f^{\mu\nu}(x;m_c,m_s),
\ee
where $d^{abc} = 2\, {\rm Tr}(\{T^a,T^b\} T^c) $, ${\rm Tr}(T^a T^b) = \frac{\delta^{ab}}{2}$
and we have indicated the colour factor explicitly. For the gauge group SU($N$), $d^{abc} d^{abc}=(N^2-1)(N^2-4)/N$
and the tensor $f^{\mu\nu}$ is the same as in Eq.\ (\ref{eq:mixed2ptQED}), to leading perturbative order.
In the low-energy EFT, i.e.\ QCD with $N_f$ quark flavours, the two-point function of Eq.\ (\ref{eq:mixed2ptQCD})
is represented by the matching coefficient times
$\< -i\partial_\lambda (\bar q(x) [\gamma^\mu,\gamma^\lambda]q(x))\; \bar q(0) \gamma^\nu q(0)\>$.
At leading non-trivial order in perturbation theory, the latter two-point function is simply $N$ times
the low-energy two-point function in the second line of Eq.\ (\ref{eq:mixed2ptQED}).
From this comparison, one obtains the colour factor and concludes that
\ba \la{eq:opDim5}
&& \bar Q \gamma^\mu Q \longrightarrow
\sum_{f=1}^{N_f} d_2(M,m_f,\alpha_s)
\,(-i)\partial_\nu(\bar q_f [\gamma^\mu,\gamma^\nu]q_f),
\nonumber\\ && d_2(M,m,z) = \frac{(N^2-1)(N^2-4)}{16 N^2}\;  d_2^{\rm QED}(M,m,z)
\stackrel{N=3}{=} \frac{5}{18}\; d_2^{\rm QED}(M,m,z).
\la{eq:d2coef}
\ea
Clearly, due to the chiral suppression, the largest contribution comes
from the heaviest quark still considered as `light'; typically, this
would be the strange quark. In terms of large-$N$ counting, $d_2$ is
of order $\lambda_H^3/N$, where $\lambda_H\equiv g_s^2 N$ is the
't Hooft coupling.  Finally, the scale at which $\alpha_s$ should be
evaluated in $d_2$ is of order $2M$.

The tensor operator has an anomalous dimension, therefore it should be evolved from the scale $2M$
to the standard renormalisation scale at which
it is defined, typically 2\,GeV in the $\overline{\rm MS}$ scheme.
The anomalous dimension is known to four-loop order in the $\overline{\rm MS}$ and in the RI$^\prime$ schemes~\cite{Gracey:2022vqr},
and the non-perturbative renormalisation of the tensor current has also been determined very recently~\cite{Chimirri:2023ovl}.
To leading order, its scale evolution reads 
\be
   {\cal O}_T(\mu) = {\cal O}_T(2M) \left(\frac{\alpha_s(\mu)}{\alpha_s(2M)} \right)^{C_F/\beta_0},
   \quad C_F = {\txts\frac{N^2-1}{2N}} , \qquad \beta_0 = {\txts\frac{11}{3}}N - {\txts\frac{2}{3}}N_f.
\ee
Since we will mostly be interested in the charm quark, for which $2M$ is not very different from $\mu=2\,$GeV,
and since we are only making order-of-magnitude estimates, we will ignore this effect in the following numerical estimates.
It should however be taken into account for making asymptotic statements on the $M$ dependence of low-energy matrix elements
of the heavy-quark current, and when estimating the effects of the bottom quark.

One can now answer questions such as `What would be the magnetic moment of the nucleon
if the photon coupled only to an asymptotically heavy quark of mass $M$?'
One finds
\be
\mu^Q_{p,n} \equiv G^Q_M(0) = F^Q_2(0) = -4 m_N \sum_{f=1}^{N_f} d_2(M,m_f,\alpha_s) \;g_{T}^{f},
\ee
where $g_{T}^f$ is the tensor charge of the nucleon with respect to flavour $f$,
\ba
&& \< N(p,s)| \bar q_f [\gamma_\mu,\gamma_\nu] q_f |N(p,s)\>
= g_{T}^f\,\bar U(p,s) [\gamma_\mu,\gamma_\nu] U(p,s).
\ea
Although the quark-mass factor gives an enhanced weight to the strange quark,
the strange tensor charge is expected to be much smaller than the sum of the up and down tensor charges.
We return to numerical estimates for the charm-current contribution in section \ref{sec:mu_c}.
A simple prediction of the considerations above is that the heavy-quark current contribution
to the magnetic moments of the hyperons occuring via the tensor currents is expected to be much larger than in the nucleon.
However, we shall see that the three-gluon operator actually dominates for the interesting cases of the charm and bottom quarks.

\section{Gluonic operators and the Euler-Heisenberg Lagrangian\la{sec:EH}}

In this section, we derive the low-energy representation of the heavy-quark current
(a) in the case that the $N_f$ light quarks are massless -- in which case the tensor current does not contribute,
or (b) in the $N_f=0$ case, i.e. for the case that the low-energy is pure gluodynamics.
The most efficient way to proceed is to write an effective Lagrangian for the coupling of the gluon fields
to an external photon. This was the method used in~\cite{Novikov:1979va} to derive the effective Lagrangian for the low-energy
$(\gamma\gamma gg)$ interaction induced by a heavy-quark loop connecting two electromagnetic vertices.
Here we are treating the case of a heavy-quark loop emanating from a single electromagnetic vertex and inducing a $\gamma ggg$ coupling.
This case was considered by Combridge~\cite{Combridge:1980sx} in the perturbative regime.
Both cases are closely related to the classic Euler-Heisenberg Lagrangian for the pure QED case~\cite{Heisenberg:1936nmg,Grozin:2009an}.
First, some notational conventions. The free fermion action is $S = \int d^4x\; \bar\psi (i \partial_\mu\gamma^\mu - m) \psi$.
The gluon action reads
\be
S^{(g)} =  -{\txts\frac{1}{2}}\int d^4x \;  \,{\rm Tr}\{G_{\mu\nu}(x) G^{\mu\nu}(x)\}
=  -{\txts\frac{1}{4}} \int d^4x \;  G_{\mu\nu}^a(x)G^{\mu\nu,a}(x) 
\ee
with  $G_{\mu\nu}(x)  = G_{\mu\nu}^a(x) T^a= (\partial_\mu A_\nu^a - \partial_\nu A_\mu^a + g f^{abc} A_\mu^b A_\nu^c)T^a$ and
$[T^a,T^b] = if^{abc} T^c$.
Following the conventions of~\cite{Peskin:1995ev}, the interactions between the gauge fields and the fermions are written
\be
S^{(Q\gamma)} + S^{(Qg)} = \int d^4x\, \Big( -e_Q \,A_\mu\, \bar Q\gamma^\mu Q + g_s\, A_\mu^a\, \bar Q \gamma^\mu\, T^a Q\Big)
\ee
with $e_Q$ the electric charge of the heavy quark, for instance $e_Q = -\frac{1}{3} |e|$ with $\alpha= e^2/(4\pi)$ for the bottom quark
and $g_s$ the (positive) strong coupling constant.
The interaction of the light quarks with the gluon field then takes the same form as in $S^{(Qg)}$.
The two possible terms in the action describing the interaction with a hypothetical photon coupling only to a heavy quark of electric charge $e_Q$ are
\ba\la{eq:EH1}
 S^{(Q\gamma)}_{\rm eff} &=& \frac{-e_Q}{2}\int d^4x \; F^{\mu\nu}(x) \bigg(
h_{1}\, {\rm Tr}\Big( G_{\mu\nu}(x)\, G_{\alpha\beta}(x) G^{\alpha\beta}(x)\Big)
\\ && \qquad \qquad\qquad\qquad  + h_{2}\, {\rm Tr}\Big({\txts\frac{1}{2}} \{ G_{\mu\alpha}(x),G_{\nu\beta}(x)\} G^{\alpha\beta}(x)\Big)\bigg),
\nonumber
\\ \la{eq:EH2}
&=& -e_Q\frac{d^{abc}}{8}\int d^4x \; F^{\mu\nu}(x) \bigg(
 h_{1} \, G_{\alpha\beta}^a(x) G^{\alpha\beta,b}(x) G_{\mu\nu}^c(x)
 + h_{2}\,  G_{\mu\alpha}^a(x) G_{\nu\beta}^b(x) G^{\alpha\beta,c}(x)\bigg) \qquad 
 \\ \la{eq:EH3}
 &=& -e_Q\int d^4x \; A^\mu(x) \;\bigg( h_{1}\partial^\nu\, {\rm Tr}\Big( G_{\mu\nu}(x)\, G_{\alpha\beta}(x) G^{\alpha\beta}(x)\Big)
 \nonumber \\ && \qquad \qquad\qquad \qquad  + h_{2} \partial^\nu\, {\rm Tr}\Big({\txts\frac{1}{2}}\{ G_{\mu\alpha}(x),G_{\nu\beta}(x)\} G^{\alpha\beta}(x)\Big)\bigg) ,
\ea
with $\{A,B\} = AB+BA$ denoting the anticommutator. The heavy-quark loop diagram enabling a coupling between a photon and three gluons,
as well as its effective representation in the low-energy effective theory, are shown in Fig.~\ref{fig:3g}.

\begin{figure}
  \centerline{\includegraphics[width=0.18\textwidth]{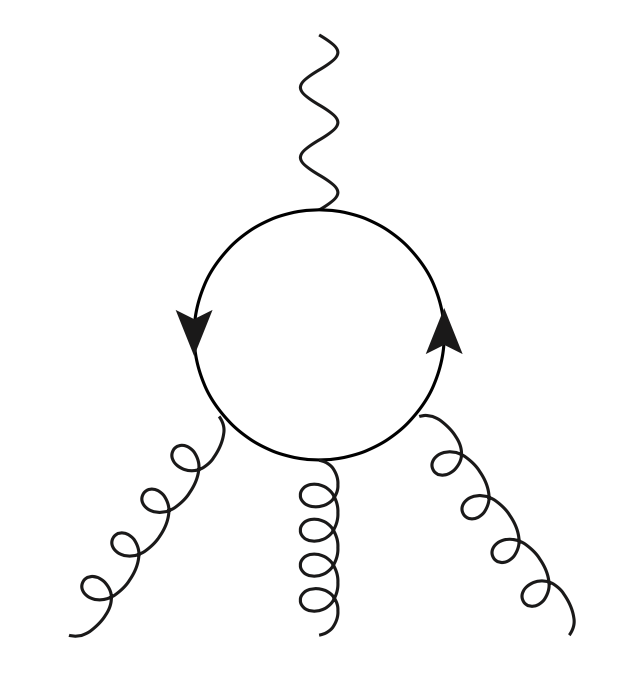}~~~~~~~~~~~~~~~~~~~~
    \includegraphics[width=0.25\textwidth]{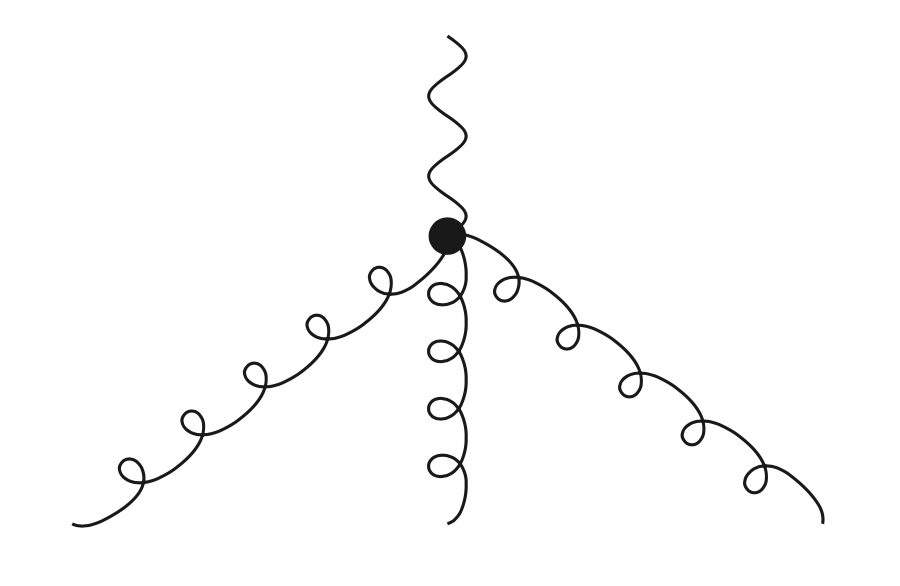}}
  \caption{\la{fig:3g} Left: the heavy-quark loop inducing a coupling between a photon and three gluons.
    Right: Effective interaction between the photon and the three gluons, described by the effective action Eq.\ (\ref{eq:EH1}).
  }
  \end{figure}

To determine the matching coefficients $h_{1}$ and $h_{2}$, 
we consider the invariant amplitude ${\cal M}$ for the scattering process of a photon and a gluon into two gluons,
\be
\gamma(p,\sigma) + g(k,\tau,a)\;\longrightarrow \;
g(p',\sigma',b) + g(k',\tau',c),
\ee
which we write ${\cal M} = -e_Q(h_{1} {\cal M}^{(1)} + h_{2} {\cal M}^{(2)})$ in the low-energy effective field theory.
The analogous $\gamma\gamma\to \gamma\gamma$ light-by-light scattering amplitudes at low energies are for instance given in the tutorial~\cite{Liang:2011sj}.
These matching coefficients can  be inferred from the well-known Euler-Heisenberg (EH) result for the pure QED case
by applying the appropriate color and combinatorial factors.
We thus start off from the EH result for the case of a heavy-lepton loop of mass $M$~\cite{Grozin:2007zz,Dunne:2004nc,Manohar:2018aog},
\ba
h^{\rm QED}_{1}(M,e_Q) &=& \frac{1}{18} \frac{e_Q^3}{16\pi^2\,M^4},
\\ h^{\rm QED}_{2}(M,e_Q) &=& -\frac{7}{45} \frac{e_Q^3}{16\pi^2\, M^4}.
\ea
Let us first inspect the one-loop calculation of the amplitude ${\cal M}$
in the complete, $N_f+1$ flavour theory with regard to
the differences between the $\gamma\gamma\to \gamma\gamma$ and $\gamma g \to gg$ cases.
First, the factor $\frac{d^{abc}}{4}$ appears in the latter case, as we have seen in section \ref{sec:tens_curr}.
The only other difference is that
 $e_Q$ must be replaced by $(-g_s)$. The sign stems from the opposite sign of the $\bar Q Q g$ interaction term in $S^{(Qg)}$
as compared to the $\bar Q Q \gamma$ term in $S^{(Q\gamma)}$.

When computing the amplitude $\gamma g \to gg$ in the low-energy effective field theory,
there are two differences with respect to the $\gamma\gamma\to\gamma\gamma$ case.
First, the factor $\frac{d^{abc}}{4}$ appears, as is clear from comparing Eq.\ (\ref{eq:EH1}) applied to QED and
Eq.\ (\ref{eq:EH2}) applied to QCD. However, precisely the same factor also appears in the calculation of the quark-loop diagram
in the $(N_f+1)$ flavour theory, as described in the previous paragraph,
so that it does not affect the matching calculation. 

Secondly, there is a combinatorial effect.
In Eq.\ (\ref{eq:EH1}) applied to QED, each field strength tensor plays the same role:
the first operator consists of two contracted field strength tensors,
the second of a single circular chain of such tensors.
Now, in the calculation of the $\gamma g \to gg$ amplitude, one specific field strength tensor
is necessarily annilating the initial-state photon, whereas in the QED case, it can be contracted with any one of the four photons
involved in the amplitude. This  results in the $\gamma\gamma\to\gamma\gamma$ amplitudes ${\cal M}^{(1)}$ and ${\cal M}^{(2)}$
being four times larger than the corresponding expressions multiplying $\frac{d^{abc}}{4}$ in the $\gamma g \to gg$ case.

The upshot is that, in order to compensate the reduced combinatorial factor,
the matching coefficients must be four times larger in the $\gamma g \to gg$ case.
We thus conclude
\ba\la{eq:h41coef}
h_{1}(M,g_s) &=& -\frac{2}{9} \frac{g_s^3}{16\pi^2\, M^4},
\\ h_{2}(M,g_s) &=& +\frac{28}{45} \frac{g_s^3}{16\pi^2\,M^4}.
\la{eq:h42coef}
\ea
In summary, using Eq.\ (\ref{eq:EH3})) we have found the following low-energy representation of the heavy-quark current,
\ba \la{eq:opDim7}
&&  \bar Q \gamma^\mu Q \longrightarrow 
 h_{1}(M,g_s)\;
\partial_\nu  {\rm Tr}\Big( G^{\mu\nu}(x)\, G_{\alpha\beta}(x) G^{\alpha\beta}(x)\Big)
\\ && \qquad \qquad +  h_{2}(M,g_s)\;
\partial_\nu {\rm Tr}\Big({\txts\frac{1}{2}} \{ G^{\mu\alpha}(x),G^{\nu\beta}(x)\} G_{\alpha\beta}(x)\Big).
\nonumber
\ea
To repeat, these two operators are parametrically the leading ones in the limit
where the light-quark masses vanish, or indeed for the $N_f=0$ (pure
gauge) theory.
In the former case, one must be aware that chirally invariant four-fermion operators such as
\[
\partial_\nu\left({\txts\sum_{f=1}^{N_f}} (\bar q_f \gamma^{[\mu} \gamma^\nu \gamma^{\lambda]} q_f) {\txts\sum_{f'=1}^{N_f}}(\bar q_{f'}  \gamma_\lambda q_{f'})\right),
\]
where the square brackets denote complete antisymmetrization of the enclosed indices,
have the same dimension as the three-gluon operators above. While their matching coefficients,
being at least of order $\alpha_s^3$, are parametrically suppressed,  
their mixing under renormalisation with the gluonic operators of Eq.\ (\ref{eq:opDim7}) should in general be taken into account.
%

Rewriting the gluonic operators in terms of the rescaled non-Abelian gauge field having
for action $S = \int d^4x (-\frac{1}{2g^2}){\rm Tr}\{{\cal G}_{\mu\nu}{\cal G}^{\mu\nu}\}$,
the computed matching coefficients formally become independent of the gauge coupling.
They are also independent of $N$.
We expect the gluonic operators ${\rm Tr}\{{\cal G}{\cal G}{\cal G}\}$ to have hadronic matrix elements of order unity,
similar to the ${\rm Tr}({\cal G}_{\mu\nu}{\cal G}_{\mu\nu})$ `trace anomaly'
operator having an order-unity matrix element on the nucleon in units of the nucleon mass~\cite{Ji:1994av}.
Thus, while the matching coefficients of the gluonic operators are suppressed by two additional powers
of $1/M$ relative to the tensor current of the light quarks, they are neither suppressed by light quark masses,
nor by a factor of $1/N$, nor by $(\alpha_s(M)/\pi)^3$.

\section{Physics applications\la{sec:appl}}

Since some of our applications are closely related to lattice QCD calculations,
we provide the low-energy representation  of the heavy-quark current in Euclidean notation.
In Euclidean space, we use hermitian Dirac matrices, $\{\gamma^{\rm _E}_\mu,\gamma^{\rm _E}_\nu\}=2\delta_{\mu\nu}$,
so that the action for a free fermion reads $S_{\rm E} = \int d^4x \; \bar \psi (\gamma^{\rm _E}_\mu\,\partial_\mu + m)\psi$,
and the Euclidean action describing the heavy quark's interactions with the gauge fields takes the form
\be
S_{\rm E}^{(Q \gamma)}+ S_{\rm E}^{(Q g)}
= \int d^4x\; \Big(-ie_Q\, A_\mu \,\bar Q\gamma^{\rm _E}_\mu\, Q +  ig\, A_\mu^a\,\bar Q\gamma^{\rm _E}_\mu\, T^a Q \Big).
\ee
We then have the operator mappings
\ba \la{eq:opDim5Eucl}
&& \bar Q \gamma^{\rm _E}_\mu Q \longrightarrow
\sum_{f=1}^{N_f} d_2(M,m_f,\alpha_s)\;
\partial_\nu(\bar q_f [\gamma^{\rm _E}_\mu,\gamma^{\rm _E}_\nu]q_f),
\ea
and 
\ba \la{eq:opDim7Eucl}
&& \bar Q \gamma^{\rm _E}_\mu Q \longrightarrow  h_{1}(M,g_s)\;
i\,\partial_\nu  {\rm Tr}\Big( G_{\mu\nu}(x)\, G_{\alpha\beta}(x) G_{\alpha\beta}(x)\Big)
\\ && \qquad \qquad +  h_{2}(M,g_s)\;
i\,\partial_\nu {\rm Tr}\Big({\txts\frac{1}{2}}\{ G_{\mu\alpha}(x),G_{\nu\beta}(x)\} G_{\alpha\beta}(x)\Big).
\nonumber
\ea
The expressions for the coefficients $d_2$, $h_{1}$ and $h_{2}$ are given in Eqs.\ (\ref{eq:d2coef}, \ref{eq:h41coef}, \ref{eq:h42coef}).

\subsection{Charm and bottom magnetic moments of the proton\la{sec:mu_c}}

In this section, we estimate the size of the charm magnetic moment of
the nucleon in units of the nuclear magneton, $\mu^c_{p,n}$.  Consider first the contribution of the
tensor currents of the light quarks in the low-energy EFT.  For the
following numerical estimates, we use the tensor charges given in
Ref.~\cite{Alexandrou:2022coj} in the $\overline{\rm MS}$ scheme at 2\,GeV, which have been computed in lattice
QCD with dynamical up, down, strange and charm quarks; see also the recent results in Ref.~\cite{Park:2023tsj}.
\ba
\mu^c_{p,n}\Big|_{\rm tensor}&=&
\frac{5}{18} ({\txts\frac{3}{2}}\zeta(3) - {\txts\frac{19}{16}}) \left(\frac{\alpha_s}{\pi}\right)^3
\frac{m_N}{m_c^2}  ({m_u}\,  g_{T}^u + {m_d}\, g_{T}^d + {m_s}\, g_{T}^s)
\\ &\simeq& 9\times 10^{-8} ({m_u}\, g_{T}^u + {m_d}\, g_{T}^d + {m_s}\, g_{T}^s)\,/\,{\rm MeV}
\simeq (3\pm 3)\times 10^{-8}.
\ea
We have used $m_c = 1.27\,{\rm GeV}$ and
the light-quark masses from the Particle Data Group~\cite{Workman:2022ynf} (which are in the $\overline{\rm MS}$ scheme at 2\,GeV),
and set $\alpha_s =0.3$. The uncertainty is large, due to a significant cancellation
between the up and down quark contributions; the strange contribution is somewhat smaller, but not negligible, due to
the $m_s$ factor. In any case, the entire tensor contribution is minuscule. The same conclusion applies \emph{a fortiori}
to the bottom magnetic moment of the nucleon, $\mu^b_{p,n}$.

We now turn to the ${\rm O}(1/m_c^4)$ contribution of the two gluonic operators.
For the second gluonic operator, which has the larger matching coefficient, we obtain a contribution on the order of 
\be
\mu^c_{p,n}\Big|_{\rm glue} =   \frac{28}{45} \frac{1}{16 \pi^2 m_c^4}\cdot (\xi\,m_N^4) = 1.1\times 10^{-3}\xi,
\ee
where $\xi$ parametrizes the forward nucleon matrix
of $\frac{1}{2}{\rm Tr}(\{{\cal G}_{\mu\alpha},{\cal G}_{\nu\beta}\}{\cal G}^{\alpha\beta})$.
We expect $|\xi|$ to be of order unity, in analogy with the matrix elements of the operator ${\rm Tr}({\cal G}_{\mu\nu}{\cal G}_{\mu\nu})$~\cite{Ji:1994av}.
Since the strangeness magnetic moment of the nucleon $\mu^s_{p,n}$ is already negative (and on the order of $-0.02$~\cite{Djukanovic:2019jtp,Alexandrou:2019olr}
with about 25\% uncertainty), one might expect the same for the contribution of heavier quarks.
The order of magnitude is consistent with the findings of Ref.\ \cite{Sufian:2020coz} in lattice QCD,
\be\la{eq:mucSufian}
\mu^c_{p,n} = (-1.27\pm 0.38_{\rm stat} \pm 0.05_{\rm syst})\times 10^{-3}.
\ee
Using the result of the latter publication, we can predict the order of magnitude of the bottom magnetic moment to be
\be
\mu^b_{p,n} \simeq \left(\frac{m_c}{m_b}\right)^4\mu^c_{p,n} \approx -1 \times 10^{-5},
\ee
while the top-quark contribution is at the level of $-4\times 10^{-12}$.
The physical magnetic moment of the proton, $\mu_p\simeq 2.793$, is known to 0.3\,ppb~\cite{Tiesinga:2021myr}.
Thus the present measurement is sensitive
to the coupling of the photon to the bottom quark, but not yet to its coupling to the top quark.

It is also worth noting that the charm contribution to the average
anomalous magnetic moment of proton and neutron, $(\mu_p + \mu_n -1)/2 \simeq -0.060$,
can be estimated as $(+2/3)$ times Eq.\ (\ref{eq:mucSufian}), yielding a one to two percent contribution.
Experimentally, the average nucleon anomalous magnetic moment is known at the
level of 3.7\,ppm~\cite{Tiesinga:2021myr}, which is still precise enough to resolve the bottom
current contribution, even taking into account the latter's small
charge factor of $-1/3$.

\subsection{Vacuum polarisation: heavy-quark contributions well below their threshold}

Consider the two-point correlation function of two quark currents in the Euclidean time-momentum representation,
\be
G^{f,f'}(x_0) = - \int d^3x\; \<(\bar q_f\gamma_1^{\rm _E} q_f)(x)\; (\bar q_{f'}\gamma_1^{\rm _E} q_{f'})(0)\>
= \frac{1}{12\pi^2} \int_0^\infty ds\, s\,R^{f,f'}(s)\,\frac{e^{-\sqrt{s}|x_0|}}{2\sqrt{s}}.
\ee
We have indicated the spectral representation of the correlator~\cite{Bernecker:2011gh}, the spectral function being normalized as the well-known $R$ ratio,
such that
\be
R(s) = \sum_{f,f'=1}^{N_f} {\cal Q}_f {\cal Q}_{f'} R^{f,f'}(s)
\ee
with ${\cal Q}_f\in\{\frac{2}{3},-\frac{1}{3}\}$ the quark charges.
We begin with the case of two distinct quark flavours $f$ and $f'$, the former being the more massive one.
In that case, the correlator receives exclusively quark-disconnected contributions.

\subsubsection{The tensor current contribution: perturbative regime}

In $G^{b,c}(x_0)$, at distances much greater than $M_\Upsilon^{-1}$, we replace the bottom current by its low-energy representation
in terms of the charm-quark tensor current according to Eq.\ (\ref{eq:opDim5}).
Due to the relatively large charm mass, no chiral suppression of this contribution takes place.
Evaluating the charm vector-tensor correlator to lowest order in perturbation theory, we obtain
\be\la{eq:Rbc}
R^{b,c}(s) =  \frac{3}{64}\,
\Big({{3}}\zeta(3) - {\txts\frac{19}{8}}\Big)\frac{(N^2-1)(N^2-4)}{N}\, \left(\frac{\alpha_s(2m_b)}{\pi}\right)^3  \frac{m_c^2}{m_b^2}
\;\sqrt{1-\frac{4m_c^2}{s}}.
\ee
In an expansion in $1/m_b^2$,
this expression is the leading perturbative contribution of the bottom current to the spectral function above the charm threshold,
but well below the bottom threshold.
We believe this result to be new. The quark-disconnected contribution has been computed for massless quarks,
in which case it is known to O($\alpha_s^4$) included~\cite{Baikov:2012er,Baikov:2012zm}.
%

\subsubsection{The tensor current contribution: hadronic regime}

Consider the correlator $G^{c,s}(x_0)$ at distances much greater than $M_{J/\psi}^{-1}$.
In this subsection, we replace the charm current by its low-energy representation
in terms of the strange-quark tensor current according to Eq.\ (\ref{eq:opDim5})
and work out an estimate for the ratio $G^{c,s}/G^{s,s}$. 

Moreover, we exploit the fact that the strange-strange correlator is dominated over
a significant interval of Euclidean times by the $\phi$ meson, of mass $M_\phi$.
This comes from the narrowness of the $\phi$ resonance ($\Gamma_\phi\simeq 4\,{\rm MeV}$),
from the suppressed coupling of the strange current to the $\omega$ meson and 
the small size of the $K\bar K$ and $\pi\pi\pi$ continua below $\sqrt{s}=1\,{\rm GeV}$.
Using the spectral decomposition, we then obtain, for $x_0$  positive and on the order of 1\,fm,
\be
G^{c,s}(x_0) \simeq d_2(m_c,m_s,\alpha_s) M_\phi \cdot  \frac{e^{-M_\phi x_0}}{2M_\phi}\,\sum_{\lambda=1}^3
\<0| \bar s [\gamma_1^{\rm _E},\gamma_0] s |\phi,\vec 0,\lambda\> \;\<\phi,\vec 0,\lambda | \bar s \gamma_1^{\rm _E} s | 0\>.
\ee
Now we insert the standard parametrization of the matrix elements of a massive vector particle,
\ba
\<0| \bar s \gamma^\mu s | \phi,\vec p,\lambda\> &=& \epsilon^\mu_{\lambda} \,f_\phi M_\phi, \phantom{\Big|}
\\
\<0| \bar s {\txts\frac{i}{2}}[\gamma^\mu,\gamma^\nu] s | \phi,\vec p,\lambda\> &=&
i\,(\epsilon^\mu_{\lambda} \,p^\nu - \epsilon^\nu_{\lambda}\, p^\mu)  \,f_\phi^\perp.
\ea
Thus we arrive at the expression
\be
G^{c,s}(x_0) \simeq  -d_2(m_c,m_s,\alpha_s)   (2f_\phi\, f_\phi^\perp\, M_\phi^3)  \frac{e^{-M_\phi x_0}}{2M_\phi},
\ee
the corresponding contribution to the spectral function reading
\be
\frac{R^{c,s}(s)}{12\pi^2} = -  \left(d_2(m_c,m_s,\alpha_s)2M_\phi\right)   (f_\phi\, f_\phi^\perp) \,\delta(s-M_\phi^2).
\ee
On the other hand, the strangeness correlator is given by
\be
G^{s,s}(x_0) 
\simeq f_\phi^2 M_\phi^2\; \frac{e^{-M_\phi x_0}}{2M_\phi} ,
\ee
corresponding to
\be
\frac{R^{s,s}(s)}{12\pi^2} =  f_\phi^2  \,\delta(s-M_\phi^2).
\ee
Taking the ratio of correlators
\ba\la{eq:GcsovGss}
\frac{G^{c,s}(x_0)}{G^{s,s}(x_0)} &\simeq& - \left(d_2(m_c,m_s,\alpha_s) 2M_\phi\right)  \frac{f_\phi^\perp}{f_\phi }
\simeq 5\cdot 10^{-6}  \cdot \frac{f_\phi^\perp}{f_\phi } \simeq 3\cdot 10^{-6}
\quad (x_0\approx 1\,{\rm fm}),
\ea
we find a very small result.
In the last step, we have assumed $f_\phi^\perp/f_\phi \approx 2/3$ based on
the lattice calculation~\cite{Braun:2016wnx}, which was for the $\rho$ meson, and references therein.
Since the ratio (\ref{eq:GcsovGss}) is very small, we turn to the role of the gluonic operators in the next subsection.
\medskip

For a long time, charmonium properties have been extracted on the lattice by neglecting the
disconnected diagram in charm-current two-point functions.
See however the dedicated study and discussion in Ref.~\cite{Levkova:2010ft},
where the effect of the disconnected diagram on the extraction of the $J/\psi$ mass could not be resolved,
in contrast to the $\eta_c$ channel. The effect of charm sea quarks, which is distinct from the one discussed here, has
recently been addressed in~\cite{Cali:2019enm}.
Here we assess the relative importance of the disconnected diagram based on representing the charm current
as a strange tensor current.
We define $G^{c,c} = G^{c,c}_{\rm conn} + G^{c,c}_{\rm disc}$ with
\ba
G^{c,c}_{\rm conn}(x_0) &=& \int d^3x\; \< {\rm Tr}\{\gamma_1 S_c(x,0) \gamma_1 S_c(0,x)\}\>,
\\
G^{c,c}_{\rm disc}(x_0) &=& -\int d^3x\; \< {\rm Tr}\{\gamma_1 S_c(x,x)\; {\rm Tr}\{\gamma_1 S_c(0,0)\}\>,
\ea
where $S_c(x,y)$ denotes the charm-quark propagator in a non-perturbative gauge field background, which is being averaged over.
At distances well beyond the inverse $J/\psi$ mass $M_\psi^{-1}$, we have
\ba
G^{c,c}_{\rm conn}(x_0) \simeq f_\psi^2 M_\psi^2 \cdot \frac{e^{-M_\psi x_0}}{2M_\psi},
\ea
whereas
\be
G^{c,c}_{\rm disc}(x_0) \simeq d_2(m_c,m_s,\alpha_s)^2\, (f_\phi^\perp)^2\;2M_\phi^4\cdot\frac{e^{-M_\phi x_0}}{2M_\phi}.
\ee
Although the matching coefficient $d_2$ is small,  $G^{c,c}_{\rm disc}$ falls off much more slowly
than $G^{c,c}_{\rm conn}(x_0)$.  With $f_\psi \simeq 0.405$\,GeV~\cite{Hatton:2020qhk} and $f_\phi^\perp \simeq 0.160$\,GeV
(an educated guess based on lattice results for $f^\perp_\rho$~\cite{Braun:2016wnx}), we reach the conclusion
that the disconnected is about 10\% of the connected at $x_0\simeq 2.4$\,fm, and of course its relative size increases
proportionally to $\exp((M_\psi - M_\phi)x_0)$ beyond that point. The effect on the effective mass (defined as the negative of the logarithmic derivative
of $G^{c,c}(x_0)$) reaches 1\% at $x_0\approx 2.2\,$fm and increases rapidly thereafter.
The $\omega$ meson also contributes, for two reasons: one is through the matching of the charm current to the light-quark tensor current,
 which is however strongly chirally suppressed, and the other is via the coupling of the strange tensor current to the $\omega$,
 whose size is unknown but presumably quite strongly Okubo-Zweig-Iizuka (OZI) suppressed.
 Asymptotically, the three-pion continuum dominates the charm correlator $G^{c,c}(x_0)$ in isospin-symmetric QCD.

The values of $x_0$ estimated above, at which the disconnected diagrams become significant, must be viewed as upper bounds,
since in the next subsection we show that the representation of the charm current via gluonic operators probably yields
a larger contribution.

\subsubsection{The contribution of the two gluonic operators}

Define the gluonic decay constants of the $\phi$ meson as follows,
\ba
\<0|   {\rm Tr}\Big( {\cal G}^{\mu\nu}(x)\, {\cal G}_{\alpha\beta}(x) {\cal G}^{\alpha\beta}(x)\Big) |\phi,\vec p,\lambda\>
&=& i\,(\epsilon^\mu_{\lambda} \,p^\nu - \epsilon^\nu_{\lambda}\, p^\mu)  \,f_\phi^{{\cal G},1} M_\phi^3,
\\
\<0| {\txts\frac{1}{2}} {\rm Tr}\Big(\{ {\cal G}^{\mu\alpha}(x),{\cal G}^{\nu\beta}(x)\} {\cal G}_{\alpha\beta}(x)\Big) |\phi,\vec p,\lambda\>
&=& i\,(\epsilon^\mu_{\lambda} \,p^\nu - \epsilon^\nu_{\lambda}\, p^\mu)  \,f_\phi^{{\cal G},2} M_\phi^3,
\ea
and similarly for the $\omega$ meson.
Based on the gluonic contribution, Eq.\ (\ref{eq:opDim7}), we then obtain
from the spectral representation, neglecting the OZI-suppressed $\omega$ meson contribution,
\ba
G^{c,s}(x_0)
\simeq \Big[ - {\txts\frac{2}{9}}  f_\phi^{{\cal G},1} 
  + {\txts\frac{28}{45}}  f^{{\cal G},2}_\phi \Big] \cdot \frac{M_\phi^5}{16 \pi^2 m_c^4}
(M_\phi f_\phi)\; \frac{e^{-M_\phi x_0}}{2M_\phi},
\ea
leading to the ratio of correlators for $x_0\approx1\,{\rm fm}$,
\be
\frac{G^{c,s}(x_0)}{G^{s,s}(x_0)} \simeq  \frac{M_\phi^4}{16 \pi^2 m_c^4}\cdot
\Big[ - {\txts\frac{2}{9}}  f_\phi^{{\cal G},1} + {\txts\frac{28}{45}}  f^{{\cal G},2}_\phi \Big] {\Big/}f_\phi
=2.6\cdot 10^{-3}\cdot \Big[ - {\txts\frac{2}{9}}  f_\phi^{{\cal G},1} + {\txts\frac{28}{45}}  f^{{\cal G},2}_\phi \Big]{\Big/}f_\phi \;.
\ee
We expect this contribution via the gluonic operators to be dominant over that of the tensor currents in Eq.\ (\ref{eq:GcsovGss})
because we see no reason why the ratios $f^{{\cal G},i}_\phi/f_\phi$ should be as small as $10^{-3}$.
By the same token, we  expect $G^{c,c}_{\rm disc}(x_0)$ to become comparable to $G^{c,c}_{\rm conn}(x_0)$ at smaller $x_0$ than
estimated in the previous subsection.
Moreover, via the gluonic operators, the $\omega$ meson contributes to $(G^{c,u}+G^{c,d})/2$ analogously to the $\phi$ meson in $G^{c,s}$,
without any chiral suppression, and both mesons contribute in a similar way to $G^{c,c}_{\rm disc}(x_0)$.

\section{Conclusion\la{sec:concl}}

We have derived the low-energy effective representation of heavy-quark
vector currents.  As a concrete perturbative result, we have obtained
the bottom-current contribution to the $R(s)$ ratio of order $(m_c^2/m_b^2)$ in the
sub-$M_\Upsilon$ region; see Eq.\ (\ref{eq:Rbc}).

The leading contributions in $1/m_Q$ to low-energy
observables associated with the set of light-quark tensor currents can
be estimated fairly reliably, but turn out to be very small in the
hadronic vacuum polarisation or the charm magnetic moment of the
nucleon. In the latter case, an existing direct lattice calculation strongly
suggests that a different mechanism is responsible for the
O($10^{-3}$) size found for $\mu^c$.  Two (non-chirally suppressed)
gluonic operators, whose matching coefficients we derived, can explain the
size of $\mu^c$ if their matrix elements are of order one in units of the nucleon mass.
From here, we estimated the size of the bottom magnetic moment of the nucleon.

Similarly in the $R(s)$ ratio, the gluonic
operators are bound to be the dominant ones in the low-energy representation of
the charm current at $s\lesssim 1\,{\rm GeV}^2$, unless the
corresponding decay constants of the $\omega$ and $\phi$ mesons turn
out to be enormously suppressed.
The correlator $(G^{c,u}+G^{c,d})$ provides a clean way to probe
the gluonic decay constants of the $\omega$ meson. 

Since the charm quark can be treated dynamically in lattice
calculations, the matching coefficients $(d_2,h_1,h_2)$ could be
determined non-perturbatively and the reliability of the low-energy
expansion directly tested.
Though technically more challenging, this type of study could also be carried out for the
axial current to test the prediction of Eq.\ (\ref{eq:axialLE}).
Finally, the operator mappings derived in this paper might also be used for the algorithmic purpose of accelerating
the stochastic calculation of disconnected charm loops over the entire lattice.

\acknowledgments{ I thank Georg von Hippel and Konstantin Ottnad for discussions
  on the charm form factors of the nucleon.
  I acknowledge the support of Deutsche Forschungsgemeinschaft (DFG) through
  the research unit FOR~5327 ``Photon-photon interactions in the Standard Model and beyond
  -- exploiting the discovery potential from MESA to the LHC'' (grant 458854507),
  and through the Cluster of Excellence ``Precision Physics, Fundamental Interactions and Structure of Matter''
  (PRISMA+ EXC 2118/1) funded within the German Excellence Strategy (project ID 39083149).
}

\bibliography{hQcurrent}

\begin{thebibliography}{35}
\expandafter\ifx\csname natexlab\endcsname\relax\def\natexlab#1{#1}\fi
\expandafter\ifx\csname bibnamefont\endcsname\relax
  \def\bibnamefont#1{#1}\fi
\expandafter\ifx\csname bibfnamefont\endcsname\relax
  \def\bibfnamefont#1{#1}\fi
\expandafter\ifx\csname citenamefont\endcsname\relax
  \def\citenamefont#1{#1}\fi
\expandafter\ifx\csname url\endcsname\relax
  \def\url#1{\texttt{#1}}\fi
\expandafter\ifx\csname urlprefix\endcsname\relax\def\urlprefix{URL }\fi
\providecommand{\bibinfo}[2]{#2}
\providecommand{\eprint}[2][]{\url{#2}}

\bibitem[{\citenamefont{Sufian et~al.}(2020)\citenamefont{Sufian, Liu,
  Alexandru, Brodsky, de~T\'eramond, Dosch, Draper, Liu, and
  Yang}}]{Sufian:2020coz}
\bibinfo{author}{\bibfnamefont{R.~S.} \bibnamefont{Sufian}},
  \bibinfo{author}{\bibfnamefont{T.}~\bibnamefont{Liu}},
  \bibinfo{author}{\bibfnamefont{A.}~\bibnamefont{Alexandru}},
  \bibinfo{author}{\bibfnamefont{S.~J.} \bibnamefont{Brodsky}},
  \bibinfo{author}{\bibfnamefont{G.~F.} \bibnamefont{de~T\'eramond}},
  \bibinfo{author}{\bibfnamefont{H.~G.} \bibnamefont{Dosch}},
  \bibinfo{author}{\bibfnamefont{T.}~\bibnamefont{Draper}},
  \bibinfo{author}{\bibfnamefont{K.-F.} \bibnamefont{Liu}}, \bibnamefont{and}
  \bibinfo{author}{\bibfnamefont{Y.-B.} \bibnamefont{Yang}},
  \bibinfo{journal}{Phys. Lett. B} \textbf{\bibinfo{volume}{808}},
  \bibinfo{pages}{135633} (\bibinfo{year}{2020}), \eprint{2003.01078}.

\bibitem[{\citenamefont{Borsanyi
  et~al.}(2018)}]{Budapest-Marseille-Wuppertal:2017okr}
\bibinfo{author}{\bibfnamefont{S.}~\bibnamefont{Borsanyi}} \bibnamefont{et~al.}
  (\bibinfo{collaboration}{Budapest-Marseille-Wuppertal}),
  \bibinfo{journal}{Phys. Rev. Lett.} \textbf{\bibinfo{volume}{121}},
  \bibinfo{pages}{022002} (\bibinfo{year}{2018}), \eprint{1711.04980}.

\bibitem[{\citenamefont{Jegerlehner}(1986)}]{Jegerlehner:1985gq}
\bibinfo{author}{\bibfnamefont{F.}~\bibnamefont{Jegerlehner}},
  \bibinfo{journal}{Z.Phys.} \textbf{\bibinfo{volume}{C32}},
  \bibinfo{pages}{195} (\bibinfo{year}{1986}).

\bibitem[{\citenamefont{C\`e et~al.}(2022)\citenamefont{C\`e, G\'erardin, von
  Hippel, Meyer, Miura, Ottnad, Risch, San~Jos\'e, Wilhelm, and
  Wittig}}]{Ce:2022eix}
\bibinfo{author}{\bibfnamefont{M.}~\bibnamefont{C\`e}},
  \bibinfo{author}{\bibfnamefont{A.}~\bibnamefont{G\'erardin}},
  \bibinfo{author}{\bibfnamefont{G.}~\bibnamefont{von Hippel}},
  \bibinfo{author}{\bibfnamefont{H.~B.} \bibnamefont{Meyer}},
  \bibinfo{author}{\bibfnamefont{K.}~\bibnamefont{Miura}},
  \bibinfo{author}{\bibfnamefont{K.}~\bibnamefont{Ottnad}},
  \bibinfo{author}{\bibfnamefont{A.}~\bibnamefont{Risch}},
  \bibinfo{author}{\bibfnamefont{T.}~\bibnamefont{San~Jos\'e}},
  \bibinfo{author}{\bibfnamefont{J.}~\bibnamefont{Wilhelm}}, \bibnamefont{and}
  \bibinfo{author}{\bibfnamefont{H.}~\bibnamefont{Wittig}},
  \bibinfo{journal}{JHEP} \textbf{\bibinfo{volume}{08}}, \bibinfo{pages}{220}
  (\bibinfo{year}{2022}), \eprint{2203.08676}.

\bibitem[{\citenamefont{Tiesinga et~al.}(2021)\citenamefont{Tiesinga, Mohr,
  Newell, and Taylor}}]{Tiesinga:2021myr}
\bibinfo{author}{\bibfnamefont{E.}~\bibnamefont{Tiesinga}},
  \bibinfo{author}{\bibfnamefont{P.~J.} \bibnamefont{Mohr}},
  \bibinfo{author}{\bibfnamefont{D.~B.} \bibnamefont{Newell}},
  \bibnamefont{and} \bibinfo{author}{\bibfnamefont{B.~N.}
  \bibnamefont{Taylor}}, \bibinfo{journal}{Rev. Mod. Phys.}
  \textbf{\bibinfo{volume}{93}}, \bibinfo{pages}{025010}
  (\bibinfo{year}{2021}).

\bibitem[{\citenamefont{Djukanovic et~al.}(2023)\citenamefont{Djukanovic, von
  Hippel, Meyer, Ottnad, Salg, and Wittig}}]{Djukanovic:2023jag}
\bibinfo{author}{\bibfnamefont{D.}~\bibnamefont{Djukanovic}},
  \bibinfo{author}{\bibfnamefont{G.}~\bibnamefont{von Hippel}},
  \bibinfo{author}{\bibfnamefont{H.~B.} \bibnamefont{Meyer}},
  \bibinfo{author}{\bibfnamefont{K.}~\bibnamefont{Ottnad}},
  \bibinfo{author}{\bibfnamefont{M.}~\bibnamefont{Salg}}, \bibnamefont{and}
  \bibinfo{author}{\bibfnamefont{H.}~\bibnamefont{Wittig}}
  (\bibinfo{year}{2023}), \eprint{2309.07491}.

\bibitem[{\citenamefont{Collins et~al.}(1978)\citenamefont{Collins, Wilczek,
  and Zee}}]{Collins:1978wz}
\bibinfo{author}{\bibfnamefont{J.~C.} \bibnamefont{Collins}},
  \bibinfo{author}{\bibfnamefont{F.}~\bibnamefont{Wilczek}}, \bibnamefont{and}
  \bibinfo{author}{\bibfnamefont{A.}~\bibnamefont{Zee}},
  \bibinfo{journal}{Phys. Rev. D} \textbf{\bibinfo{volume}{18}},
  \bibinfo{pages}{242} (\bibinfo{year}{1978}).

\bibitem[{\citenamefont{Chetyrkin and Kuhn}(1993)}]{Chetyrkin:1993hk}
\bibinfo{author}{\bibfnamefont{K.~G.} \bibnamefont{Chetyrkin}}
  \bibnamefont{and} \bibinfo{author}{\bibfnamefont{J.~H.} \bibnamefont{Kuhn}},
  \bibinfo{journal}{Z. Phys. C} \textbf{\bibinfo{volume}{60}},
  \bibinfo{pages}{497} (\bibinfo{year}{1993}).

\bibitem[{\citenamefont{Shifman et~al.}(1978)\citenamefont{Shifman, Vainshtein,
  and Zakharov}}]{Shifman:1978zn}
\bibinfo{author}{\bibfnamefont{M.~A.} \bibnamefont{Shifman}},
  \bibinfo{author}{\bibfnamefont{A.~I.} \bibnamefont{Vainshtein}},
  \bibnamefont{and} \bibinfo{author}{\bibfnamefont{V.~I.}
  \bibnamefont{Zakharov}}, \bibinfo{journal}{Phys. Lett. B}
  \textbf{\bibinfo{volume}{78}}, \bibinfo{pages}{443} (\bibinfo{year}{1978}).

\bibitem[{\citenamefont{Grozin}(2009)}]{Grozin:2009an}
\bibinfo{author}{\bibfnamefont{A.~G.} \bibnamefont{Grozin}}, in
  \emph{\bibinfo{booktitle}{{Helmholtz International School - Workshop on
  Calculations for Modern and Future Colliders}}} (\bibinfo{year}{2009}),
  \eprint{0908.4392}.

\bibitem[{\citenamefont{Laporta and Remiddi}(1993)}]{Laporta:1992pa}
\bibinfo{author}{\bibfnamefont{S.}~\bibnamefont{Laporta}} \bibnamefont{and}
  \bibinfo{author}{\bibfnamefont{E.}~\bibnamefont{Remiddi}},
  \bibinfo{journal}{Phys. Lett. B} \textbf{\bibinfo{volume}{301}},
  \bibinfo{pages}{440} (\bibinfo{year}{1993}).

\bibitem[{\citenamefont{Kuhn et~al.}(2003)\citenamefont{Kuhn, Onishchenko,
  Pivovarov, and Veretin}}]{Kuhn:2003pu}
\bibinfo{author}{\bibfnamefont{J.~H.} \bibnamefont{Kuhn}},
  \bibinfo{author}{\bibfnamefont{A.~I.} \bibnamefont{Onishchenko}},
  \bibinfo{author}{\bibfnamefont{A.~A.} \bibnamefont{Pivovarov}},
  \bibnamefont{and} \bibinfo{author}{\bibfnamefont{O.~L.}
  \bibnamefont{Veretin}}, \bibinfo{journal}{Phys. Rev. D}
  \textbf{\bibinfo{volume}{68}}, \bibinfo{pages}{033018}
  (\bibinfo{year}{2003}), \eprint{hep-ph/0301151}.

\bibitem[{\citenamefont{Gracey}(2022)}]{Gracey:2022vqr}
\bibinfo{author}{\bibfnamefont{J.~A.} \bibnamefont{Gracey}},
  \bibinfo{journal}{Phys. Rev. D} \textbf{\bibinfo{volume}{106}},
  \bibinfo{pages}{085008} (\bibinfo{year}{2022}), \eprint{2208.14527}.

\bibitem[{\citenamefont{Chimirri et~al.}(2023)\citenamefont{Chimirri, Fritzsch,
  Heitger, Joswig, Panero, Pena, and Preti}}]{Chimirri:2023ovl}
\bibinfo{author}{\bibfnamefont{L.}~\bibnamefont{Chimirri}},
  \bibinfo{author}{\bibfnamefont{P.}~\bibnamefont{Fritzsch}},
  \bibinfo{author}{\bibfnamefont{J.}~\bibnamefont{Heitger}},
  \bibinfo{author}{\bibfnamefont{F.}~\bibnamefont{Joswig}},
  \bibinfo{author}{\bibfnamefont{M.}~\bibnamefont{Panero}},
  \bibinfo{author}{\bibfnamefont{C.}~\bibnamefont{Pena}}, \bibnamefont{and}
  \bibinfo{author}{\bibfnamefont{D.}~\bibnamefont{Preti}}
  (\bibinfo{year}{2023}), \eprint{2309.04314}.

\bibitem[{\citenamefont{Novikov et~al.}(1980)\citenamefont{Novikov, Shifman,
  Vainshtein, and Zakharov}}]{Novikov:1979va}
\bibinfo{author}{\bibfnamefont{V.~A.} \bibnamefont{Novikov}},
  \bibinfo{author}{\bibfnamefont{M.~A.} \bibnamefont{Shifman}},
  \bibinfo{author}{\bibfnamefont{A.~I.} \bibnamefont{Vainshtein}},
  \bibnamefont{and} \bibinfo{author}{\bibfnamefont{V.~I.}
  \bibnamefont{Zakharov}}, \bibinfo{journal}{Nucl. Phys.}
  \textbf{\bibinfo{volume}{B165}}, \bibinfo{pages}{67} (\bibinfo{year}{1980}).

\bibitem[{\citenamefont{Combridge}(1980)}]{Combridge:1980sx}
\bibinfo{author}{\bibfnamefont{B.~L.} \bibnamefont{Combridge}},
  \bibinfo{journal}{Nucl. Phys. B} \textbf{\bibinfo{volume}{174}},
  \bibinfo{pages}{243} (\bibinfo{year}{1980}).

\bibitem[{\citenamefont{Heisenberg and Euler}(1936)}]{Heisenberg:1936nmg}
\bibinfo{author}{\bibfnamefont{W.}~\bibnamefont{Heisenberg}} \bibnamefont{and}
  \bibinfo{author}{\bibfnamefont{H.}~\bibnamefont{Euler}}, \bibinfo{journal}{Z.
  Phys.} \textbf{\bibinfo{volume}{98}}, \bibinfo{pages}{714}
  (\bibinfo{year}{1936}), \eprint{physics/0605038}.

\bibitem[{\citenamefont{Peskin and Schroeder}(1995)}]{Peskin:1995ev}
\bibinfo{author}{\bibfnamefont{M.~E.} \bibnamefont{Peskin}} \bibnamefont{and}
  \bibinfo{author}{\bibfnamefont{D.~V.} \bibnamefont{Schroeder}},
  \emph{\bibinfo{title}{{An Introduction to quantum field theory}}}
  (\bibinfo{publisher}{Addison-Wesley}, \bibinfo{address}{Reading, USA},
  \bibinfo{year}{1995}), ISBN \bibinfo{isbn}{978-0-201-50397-5}.

\bibitem[{\citenamefont{Liang and Czarnecki}(2012)}]{Liang:2011sj}
\bibinfo{author}{\bibfnamefont{Y.}~\bibnamefont{Liang}} \bibnamefont{and}
  \bibinfo{author}{\bibfnamefont{A.}~\bibnamefont{Czarnecki}},
  \bibinfo{journal}{Can. J. Phys.} \textbf{\bibinfo{volume}{90}},
  \bibinfo{pages}{11} (\bibinfo{year}{2012}), \eprint{1111.6126}.

\bibitem[{\citenamefont{Grozin}(2007)}]{Grozin:2007zz}
\bibinfo{author}{\bibfnamefont{A.}~\bibnamefont{Grozin}},
  \emph{\bibinfo{title}{{Lectures on QED and QCD: Practical calculation and
  renormalization of one- and multi-loop Feynman diagrams}}}
  (\bibinfo{publisher}{World Scientific}, \bibinfo{address}{Singapore},
  \bibinfo{year}{2007}), ISBN \bibinfo{isbn}{978-981-256-914-1,
  978-981-256-914-1}.

\bibitem[{\citenamefont{Dunne}(2004)}]{Dunne:2004nc}
\bibinfo{author}{\bibfnamefont{G.~V.} \bibnamefont{Dunne}},
  \emph{\bibinfo{title}{{Heisenberg-Euler effective Lagrangians: Basics and
  extensions}}} (\bibinfo{year}{2004}), pp. \bibinfo{pages}{445--522},
  \eprint{hep-th/0406216}.

\bibitem[{\citenamefont{Manohar}(2018)}]{Manohar:2018aog}
\bibinfo{author}{\bibfnamefont{A.~V.} \bibnamefont{Manohar}}
  (\bibinfo{year}{2018}), \eprint{1804.05863}.

\bibitem[{\citenamefont{Ji}(1995)}]{Ji:1994av}
\bibinfo{author}{\bibfnamefont{X.-D.} \bibnamefont{Ji}},
  \bibinfo{journal}{Phys. Rev. Lett.} \textbf{\bibinfo{volume}{74}},
  \bibinfo{pages}{1071} (\bibinfo{year}{1995}), \eprint{hep-ph/9410274}.

\bibitem[{\citenamefont{Alexandrou et~al.}(2023)}]{Alexandrou:2022coj}
\bibinfo{author}{\bibfnamefont{C.}~\bibnamefont{Alexandrou}}
  \bibnamefont{et~al.}, \bibinfo{journal}{PoS}
  \textbf{\bibinfo{volume}{LATTICE2022}}, \bibinfo{pages}{092}
  (\bibinfo{year}{2023}), \eprint{2210.05743}.

\bibitem[{\citenamefont{Park et~al.}(2023)\citenamefont{Park, Bhattacharya,
  Gupta, Lin, Mondal, and Yoon}}]{Park:2023tsj}
\bibinfo{author}{\bibfnamefont{S.}~\bibnamefont{Park}},
  \bibinfo{author}{\bibfnamefont{T.}~\bibnamefont{Bhattacharya}},
  \bibinfo{author}{\bibfnamefont{R.}~\bibnamefont{Gupta}},
  \bibinfo{author}{\bibfnamefont{H.-W.} \bibnamefont{Lin}},
  \bibinfo{author}{\bibfnamefont{S.}~\bibnamefont{Mondal}}, \bibnamefont{and}
  \bibinfo{author}{\bibfnamefont{B.}~\bibnamefont{Yoon}},
  \bibinfo{journal}{PoS} \textbf{\bibinfo{volume}{LATTICE2022}},
  \bibinfo{pages}{118} (\bibinfo{year}{2023}), \eprint{2301.07890}.

\bibitem[{\citenamefont{Workman et~al.}(2022)}]{Workman:2022ynf}
\bibinfo{author}{\bibfnamefont{R.~L.} \bibnamefont{Workman}}
  \bibnamefont{et~al.} (\bibinfo{collaboration}{Particle Data Group}),
  \bibinfo{journal}{PTEP} \textbf{\bibinfo{volume}{2022}},
  \bibinfo{pages}{083C01} (\bibinfo{year}{2022}).

\bibitem[{\citenamefont{Djukanovic et~al.}(2019)\citenamefont{Djukanovic,
  Ottnad, Wilhelm, and Wittig}}]{Djukanovic:2019jtp}
\bibinfo{author}{\bibfnamefont{D.}~\bibnamefont{Djukanovic}},
  \bibinfo{author}{\bibfnamefont{K.}~\bibnamefont{Ottnad}},
  \bibinfo{author}{\bibfnamefont{J.}~\bibnamefont{Wilhelm}}, \bibnamefont{and}
  \bibinfo{author}{\bibfnamefont{H.}~\bibnamefont{Wittig}},
  \bibinfo{journal}{Phys. Rev. Lett.} \textbf{\bibinfo{volume}{123}},
  \bibinfo{pages}{212001} (\bibinfo{year}{2019}), \eprint{1903.12566}.

\bibitem[{\citenamefont{Alexandrou et~al.}(2020)\citenamefont{Alexandrou,
  Bacchio, Constantinou, Finkenrath, Hadjiyiannakou, Jansen, and
  Koutsou}}]{Alexandrou:2019olr}
\bibinfo{author}{\bibfnamefont{C.}~\bibnamefont{Alexandrou}},
  \bibinfo{author}{\bibfnamefont{S.}~\bibnamefont{Bacchio}},
  \bibinfo{author}{\bibfnamefont{M.}~\bibnamefont{Constantinou}},
  \bibinfo{author}{\bibfnamefont{J.}~\bibnamefont{Finkenrath}},
  \bibinfo{author}{\bibfnamefont{K.}~\bibnamefont{Hadjiyiannakou}},
  \bibinfo{author}{\bibfnamefont{K.}~\bibnamefont{Jansen}}, \bibnamefont{and}
  \bibinfo{author}{\bibfnamefont{G.}~\bibnamefont{Koutsou}},
  \bibinfo{journal}{Phys. Rev. D} \textbf{\bibinfo{volume}{101}},
  \bibinfo{pages}{031501} (\bibinfo{year}{2020}), \eprint{1909.10744}.

\bibitem[{\citenamefont{Bernecker and Meyer}(2011)}]{Bernecker:2011gh}
\bibinfo{author}{\bibfnamefont{D.}~\bibnamefont{Bernecker}} \bibnamefont{and}
  \bibinfo{author}{\bibfnamefont{H.~B.} \bibnamefont{Meyer}},
  \bibinfo{journal}{Eur.Phys.J.} \textbf{\bibinfo{volume}{A47}},
  \bibinfo{pages}{148} (\bibinfo{year}{2011}), \eprint{1107.4388}.

\bibitem[{\citenamefont{Baikov et~al.}(2012{\natexlab{a}})\citenamefont{Baikov,
  Chetyrkin, Kuhn, and Rittinger}}]{Baikov:2012er}
\bibinfo{author}{\bibfnamefont{P.~A.} \bibnamefont{Baikov}},
  \bibinfo{author}{\bibfnamefont{K.~G.} \bibnamefont{Chetyrkin}},
  \bibinfo{author}{\bibfnamefont{J.~H.} \bibnamefont{Kuhn}}, \bibnamefont{and}
  \bibinfo{author}{\bibfnamefont{J.}~\bibnamefont{Rittinger}},
  \bibinfo{journal}{Phys. Rev. Lett.} \textbf{\bibinfo{volume}{108}},
  \bibinfo{pages}{222003} (\bibinfo{year}{2012}{\natexlab{a}}),
  \eprint{1201.5804}.

\bibitem[{\citenamefont{Baikov et~al.}(2012{\natexlab{b}})\citenamefont{Baikov,
  Chetyrkin, Kuhn, and Rittinger}}]{Baikov:2012zm}
\bibinfo{author}{\bibfnamefont{P.}~\bibnamefont{Baikov}},
  \bibinfo{author}{\bibfnamefont{K.}~\bibnamefont{Chetyrkin}},
  \bibinfo{author}{\bibfnamefont{J.}~\bibnamefont{Kuhn}}, \bibnamefont{and}
  \bibinfo{author}{\bibfnamefont{J.}~\bibnamefont{Rittinger}},
  \bibinfo{journal}{JHEP} \textbf{\bibinfo{volume}{1207}}, \bibinfo{pages}{017}
  (\bibinfo{year}{2012}{\natexlab{b}}), \eprint{1206.1284}.

\bibitem[{\citenamefont{Braun et~al.}(2017)}]{Braun:2016wnx}
\bibinfo{author}{\bibfnamefont{V.~M.} \bibnamefont{Braun}}
  \bibnamefont{et~al.}, \bibinfo{journal}{JHEP} \textbf{\bibinfo{volume}{04}},
  \bibinfo{pages}{082} (\bibinfo{year}{2017}), \eprint{1612.02955}.

\bibitem[{\citenamefont{Levkova and DeTar}(2011)}]{Levkova:2010ft}
\bibinfo{author}{\bibfnamefont{L.}~\bibnamefont{Levkova}} \bibnamefont{and}
  \bibinfo{author}{\bibfnamefont{C.}~\bibnamefont{DeTar}},
  \bibinfo{journal}{Phys. Rev. D} \textbf{\bibinfo{volume}{83}},
  \bibinfo{pages}{074504} (\bibinfo{year}{2011}), \eprint{1012.1837}.

\bibitem[{\citenamefont{Cali et~al.}(2019)\citenamefont{Cali, Knechtli, and
  Korzec}}]{Cali:2019enm}
\bibinfo{author}{\bibfnamefont{S.}~\bibnamefont{Cali}},
  \bibinfo{author}{\bibfnamefont{F.}~\bibnamefont{Knechtli}}, \bibnamefont{and}
  \bibinfo{author}{\bibfnamefont{T.}~\bibnamefont{Korzec}},
  \bibinfo{journal}{Eur. Phys. J. C} \textbf{\bibinfo{volume}{79}},
  \bibinfo{pages}{607} (\bibinfo{year}{2019}), \eprint{1905.12971}.

\bibitem[{\citenamefont{Hatton et~al.}(2020)\citenamefont{Hatton, Davies,
  Galloway, Koponen, Lepage, and Lytle}}]{Hatton:2020qhk}
\bibinfo{author}{\bibfnamefont{D.}~\bibnamefont{Hatton}},
  \bibinfo{author}{\bibfnamefont{C.~T.~H.} \bibnamefont{Davies}},
  \bibinfo{author}{\bibfnamefont{B.}~\bibnamefont{Galloway}},
  \bibinfo{author}{\bibfnamefont{J.}~\bibnamefont{Koponen}},
  \bibinfo{author}{\bibfnamefont{G.~P.} \bibnamefont{Lepage}},
  \bibnamefont{and} \bibinfo{author}{\bibfnamefont{A.~T.} \bibnamefont{Lytle}}
  (\bibinfo{collaboration}{HPQCD}), \bibinfo{journal}{Phys. Rev. D}
  \textbf{\bibinfo{volume}{102}}, \bibinfo{pages}{054511}
  (\bibinfo{year}{2020}), \eprint{2005.01845}.

\end{thebibliography}

\end{document}